\documentstyle[twoside,fleqn,espcrc2,epsf]{article}

\newcommand{\AmS}{{\protect\the\textfont2
  A\kern-.1667em\lower.5ex\hbox{M}\kern-.125emS}}

 \title{ Visualization of topological objects in QCD
 \thanks{Supported
        in part by FWF under Contract No.~P11456}}

\author{Markus 
Feurstein\addtocounter{address}{-1}\address{Institut f\"ur Kernphysik,
Technische Universit\"at Wien, A--1040 Vienna, Austria},
Harald Markum and Stefan Thurner}

\begin{document}

\begin{abstract}
 Recently evidence appeared that instantons and monopoles
have a certain local correlation in four-dimensional pure $SU(2)$ and 
$SU(3)$ gauge theory. We visualize several specific gauge field configurations 
and show directly that there is an enhanced probability for finding 
maximally projected abelian 
monopole loops in the vicinity of instantons. This feature is independent 
of the topological charge definition used. 
\end{abstract}
\maketitle

{\it 1. Introduction and Theory.}
Two different kinds
of topological objects  seem to be important for the confinement mechanism:
color magnetic monopoles and instantons.
Color magnetic monopoles play the essential role in the dual
superconductor hypothesis. 
Instantons are supposed to cause confinement
in QCD if they form a so-called instanton liquid.
In a series of papers we have 
presented evidence that there exist local correlations between these 
topological objects on the basis of gauge field averages \cite{wir}. 
In this contribution we present a graphical approach to this subject 
by visualization of specific $SU(2)$ gluon configurations. 

In order to investigate monopole currents one has to project $SU(2)$
onto its  abelian degrees of freedom, such that an abelian $U(1) $ 
theory remains \cite{thooft2}. This  can be achieved by
various gauge fixing procedures. We employ the
so-called maximum abelian gauge which is most favorable for our
purposes. 
For the definition of the monopole currents $m(x,\mu)$ we use the 
standard method \cite{SCH87}. 
The magnetic currents form closed loops on the dual lattice as a consequence
of monopole current conservation.
From the monopole currents we define the local monopole density as 
$ \rho(x) = \frac{1}{ 4V_{4}} \sum_{\mu} | m(x,\mu) | \ .  $
%
For the implementation of the topological charge on the lattice
we use both the field theoretic plaquette and hypercube definitions 
\cite{divecchia} and a geometric definition suggested by L\"uscher 
\cite{LUE85}. All  types of topological charges 
employed are locally gauge invariant in contrary to  the 
monopole currents. 

To identify the instantons as classical solutions of the 
equations of motion on the lattice, we  cool the gauge fields. 
The cooling procedure systematically reduces quantum fluctuations 
and suppresses
differences between the different definitions of the topological charge.
In our investigation we have used  the so-called 
``Cabbibo-Marinari method''.

{\it 2. Results.}
Our configurations were produced  on a $12^{3} \times 4$ lattice with
periodic boundary conditions using the Metropolis algorithm.
The topological observables were studied in pure $SU(2)$ in 
the confinement phase at inverse gauge coupling
$\beta=4/g^{2}=2.25$, employing the
Wilson plaquette action for the gluons. 
Each of these configurations was first cooled and then
subjected to 300 gauge fixing steps enforcing the maximum abelian gauge.
Altogether 100 independent 
configurations have been produced.

For the purpose of displaying instantons  it is necessary 
to estimate their average size. We therefore computed the auto-correlation 
function of  the topological 
charge density.  
In Fig. 1 we present the results for the hypercube  and the L\"uscher 
method  after 20 cooling sweeps. 
The solid curves are fits obtained from  a convolution 
$f(x)=\int Q(t)Q(x-t)\, dt$ 
of the topological charge density 
$ Q_\sigma(x)= \frac{6}{\pi^2 \sigma^4}\,\, (\frac{\sigma^2}{x^2 +
\sigma^2})^4$
for a single instanton of size $\sigma$.
Such a fit is justified if instantons are well separated which is the 
case after 20 cooling sweeps. In lattice units the sizes of the 
hypercube and  plaquette instantons turn out as  $\sigma \sim 2.1$, the size of 
the L\"uscher instanton is $\sigma \sim 1.0$. 
\clearpage
\begin{figure}[h]
\begin{center}
\hspace{1.2cm} Instanton-Instanton Correlation
\vspace{0.2cm}
\epsfxsize=7.5cm\epsffile{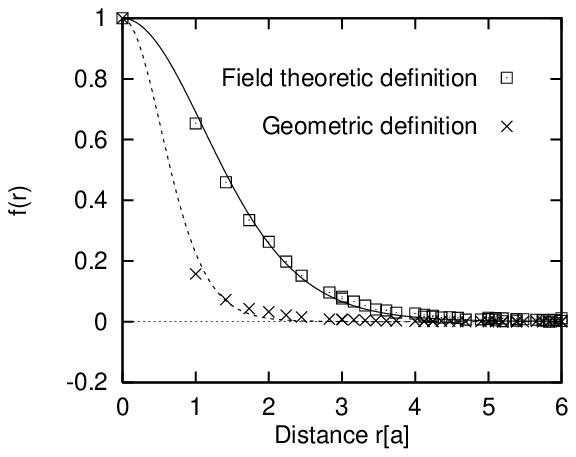} 
\\
\end{center}
{\baselineskip=13pt
Figure 1.~Auto-correlation functions of the topological charge density
after $20$ cooling sweeps. 
The curves 
represent fits to a convolution of the topological charge density for 
a single instanton of size $\sigma$. 
The L\"uscher instanton  is about half the size of the field theoretic 
definitions which yield $\sigma=2.1$. 
\baselineskip=15pt}
\vspace{0.1cm}\\
\begin{center}
Single instanton solution \vspace{-0.8cm}\\
\hspace{10cm}\epsfxsize=6.5cm\epsffile{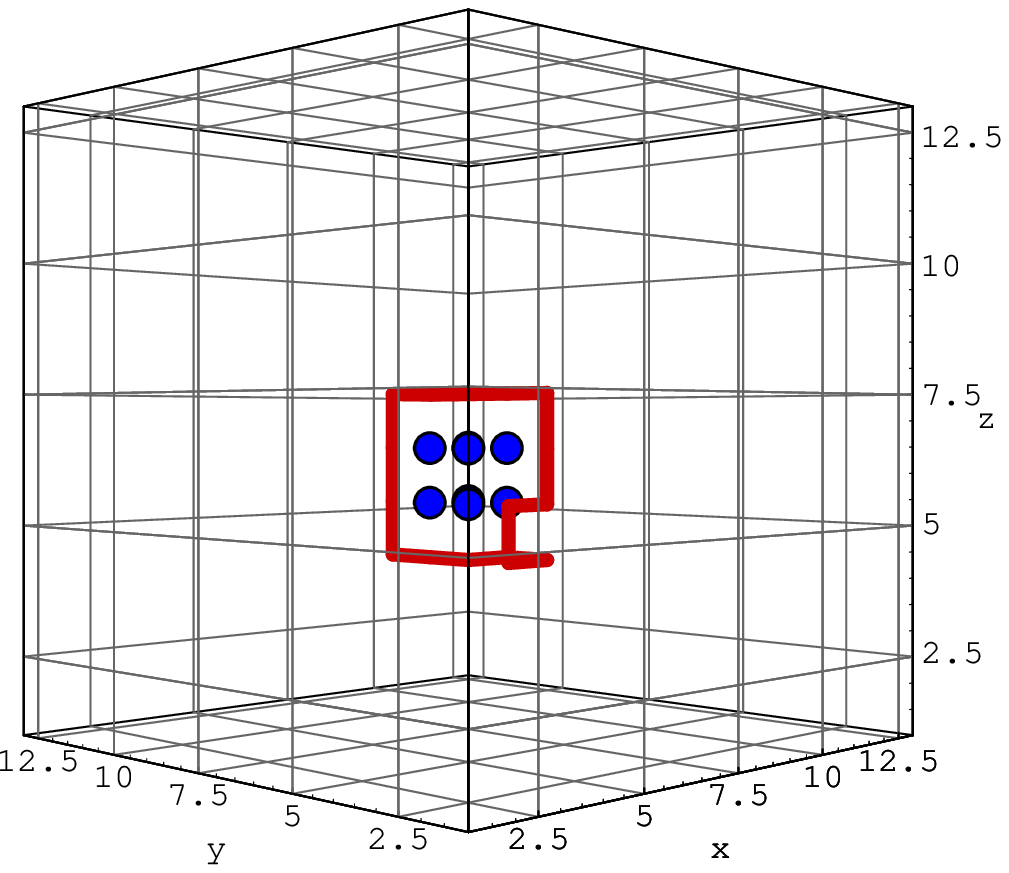} 
\\
\end{center}
{\baselineskip=13pt
Figure 2.~The location of a single instanton (dots) at constant time 
which was put on a trivial gauge field configuration is shown. 
A closed monopole loop (line) runs around the instanton. 
\baselineskip=15pt}
\end{figure}
\newpage 

\begin{figure}[b]
\begin{center}
\vspace{-1cm}
\begin{tabular}{c}
Plaquette definition \vspace{-0.3cm}\\
\epsfxsize=5.0cm\epsffile{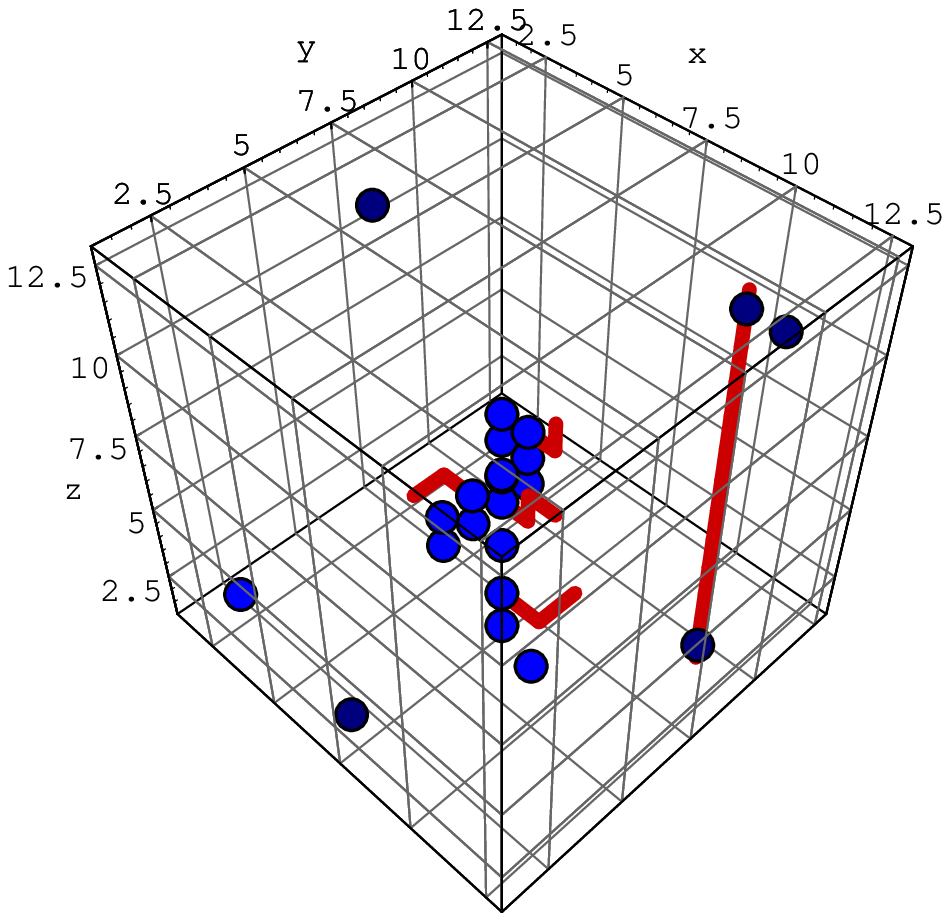}
\\ \\
Hypercube definition \vspace{-0.3cm}\\
\epsfxsize=5.0cm\epsffile{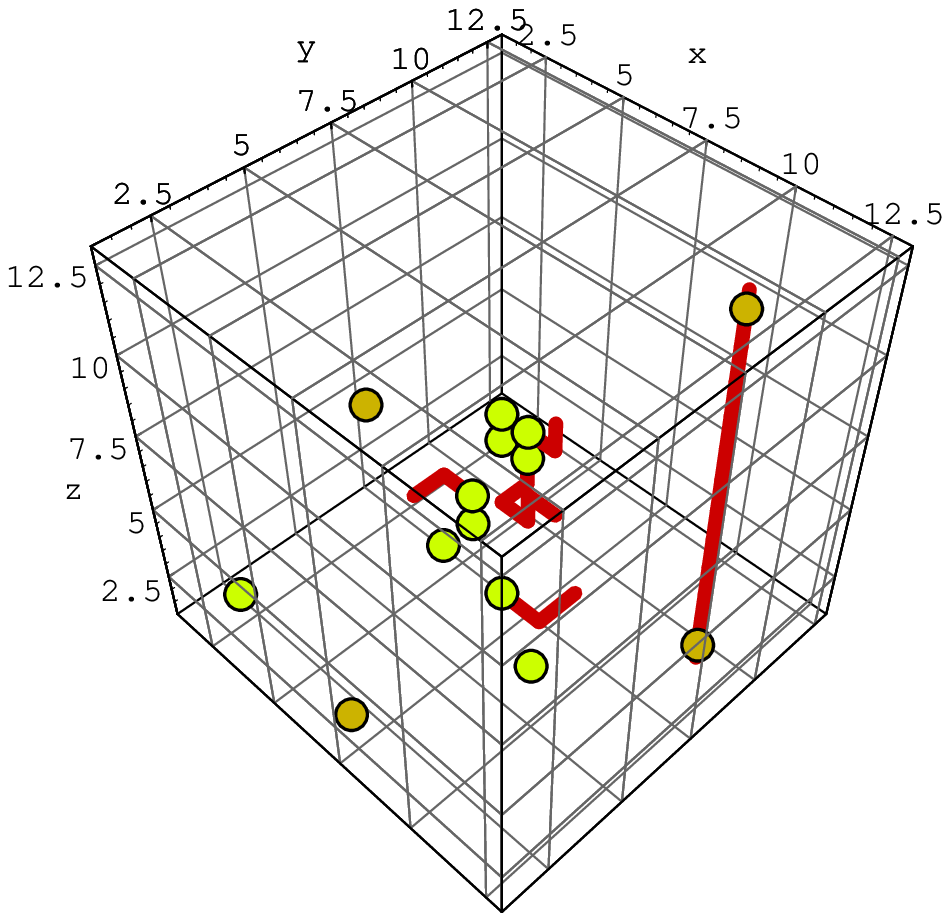}
\\ \\
L\"uscher definition \vspace{-0.3cm}\\
\epsfxsize=5.0cm\epsffile{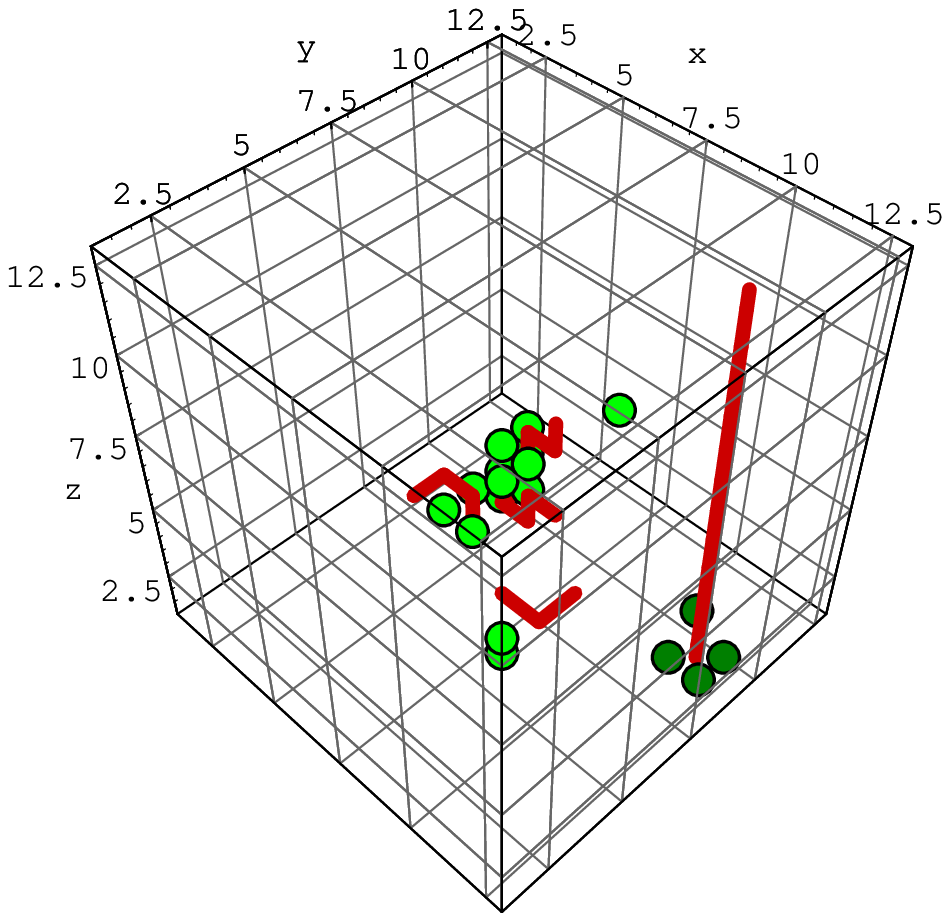}
\\ 
\end{tabular}
\end{center}
{\baselineskip=13pt
Figure 3.~Different definitions of the topological  charge 
for a specific gluon configuration after $20$ 
cooling sweeps. The instantons reside at the same places for all definitions 
and are surrounded by monopoles. 
In this particular configuration a monopole loop wraps around the torus.
\baselineskip=15pt}
\end{figure}
\clearpage 
\begin{figure*}[t]
\begin{center}
\begin{tabular}{ccc}
\vspace{-0.5cm}\\
0 cooling sweeps & 5 cooling sweeps & 11 cooling sweeps \vspace{-0.2cm}\\
\epsfxsize=5.0cm\epsffile{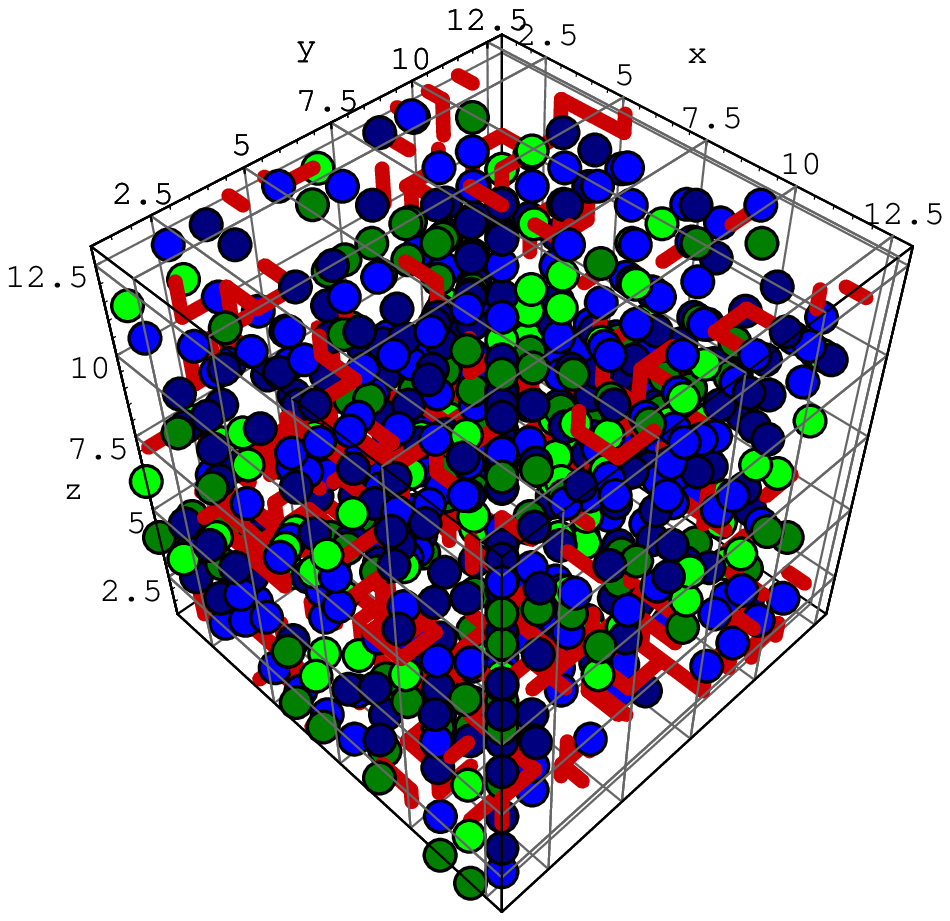}
&
\epsfxsize=5.0cm\epsffile{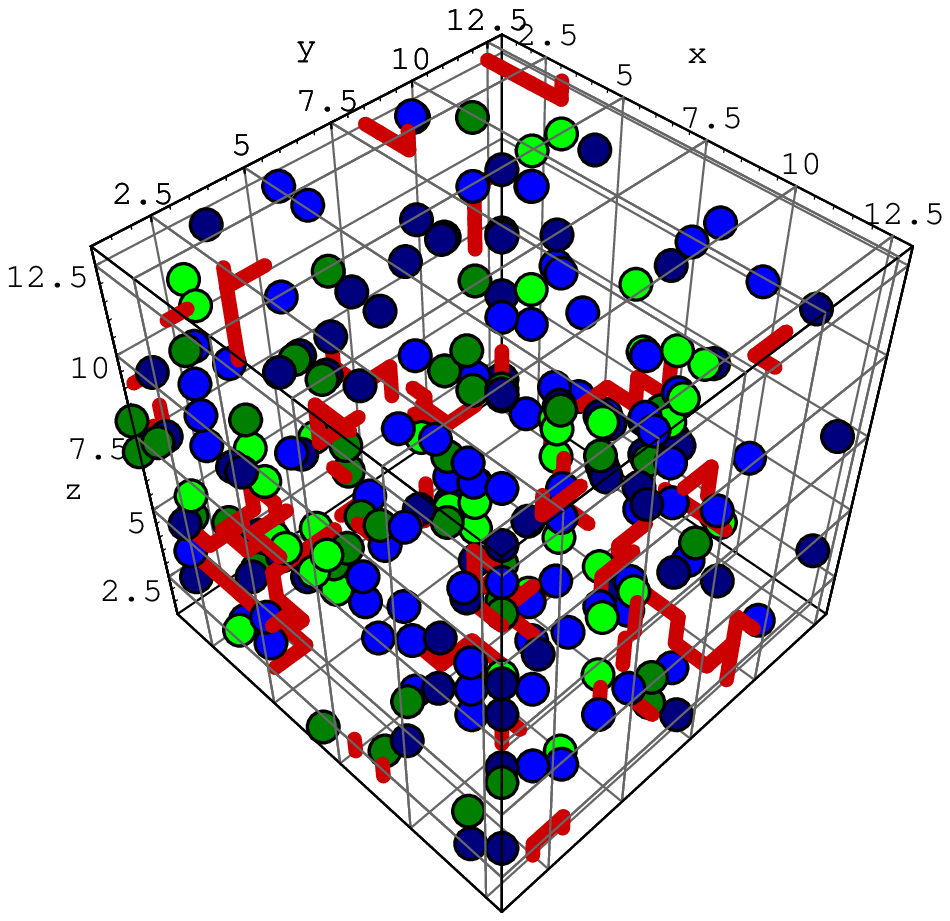} 
& 
\epsfxsize=5.0cm\epsffile{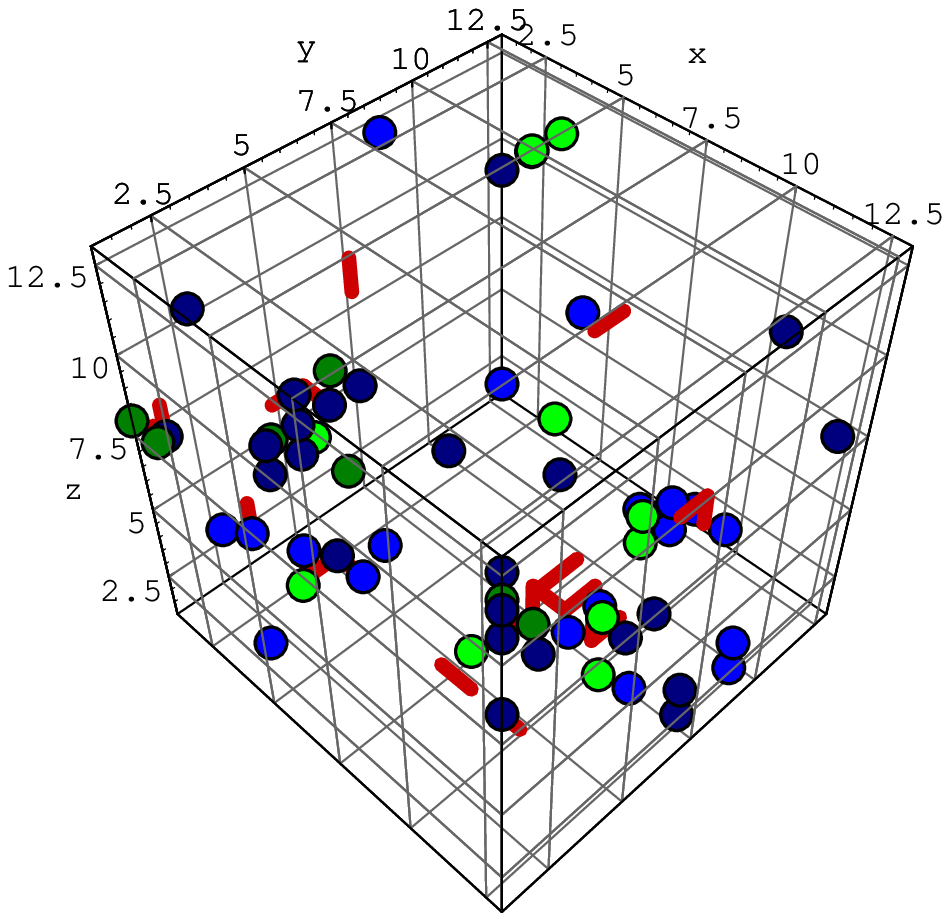}
\vspace{0.6cm}\\ 
15 cooling sweeps & 21 cooling sweeps & 25 cooling sweeps \vspace{-0.2cm}\\
\epsfxsize=5.0cm\epsffile{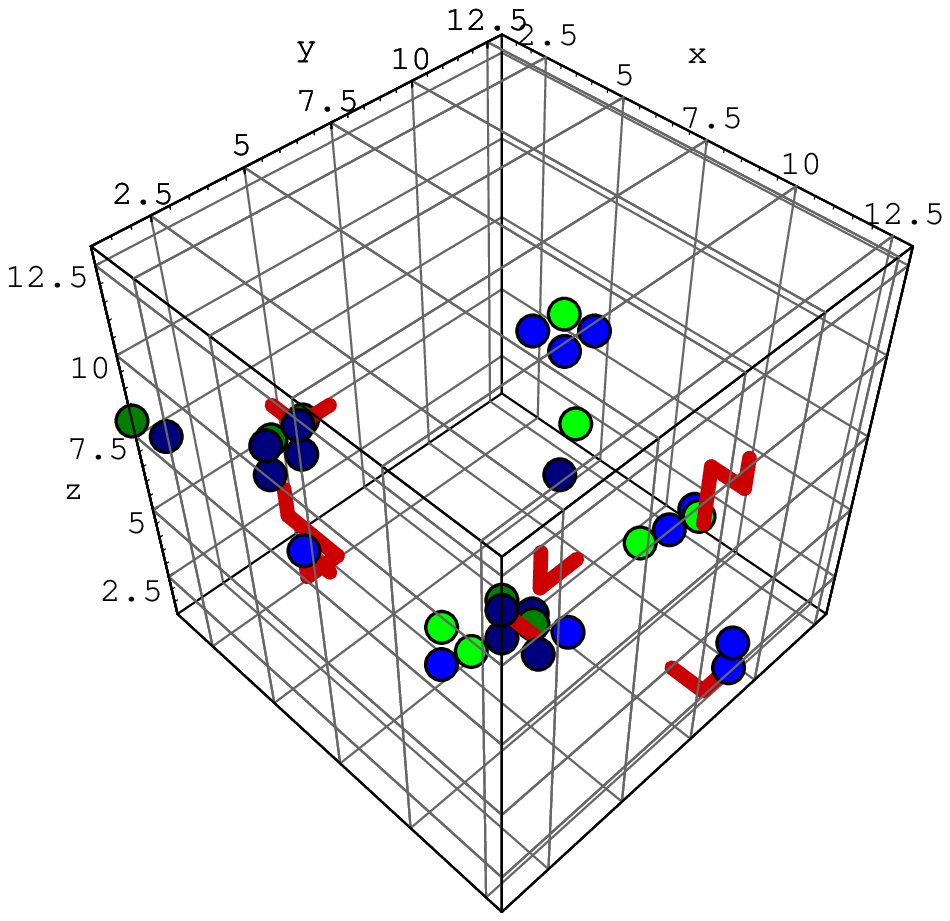}
&
\epsfxsize=5.0cm\epsffile{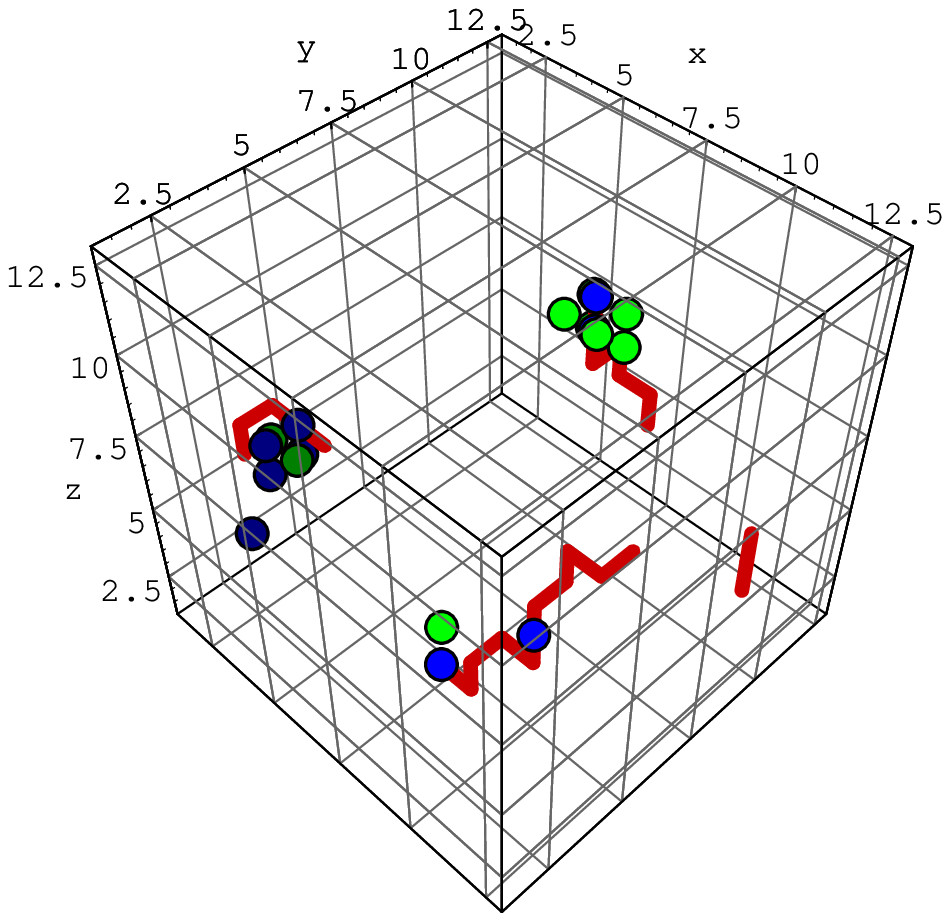} 
&
\epsfxsize=5.0cm\epsffile{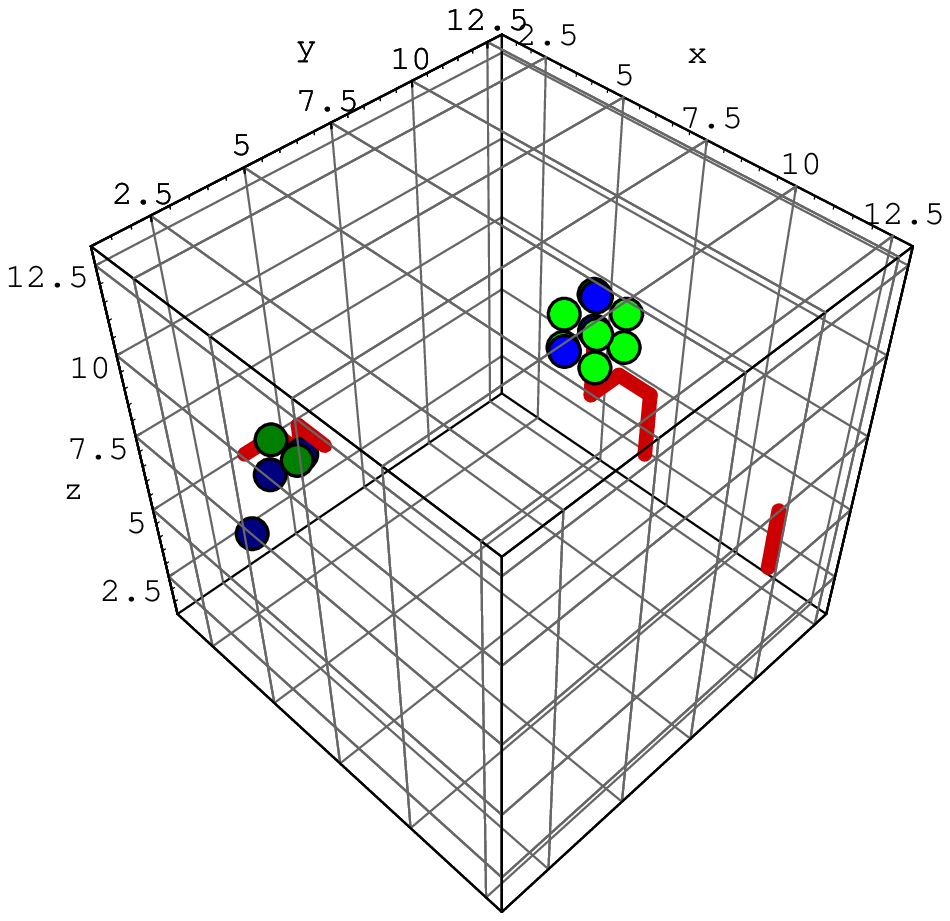}
\vspace{0.6cm}\\ 
30 cooling sweeps & 35 cooling sweeps & 40 cooling sweeps \vspace{-0.2cm}\\
\epsfxsize=5.0cm\epsffile{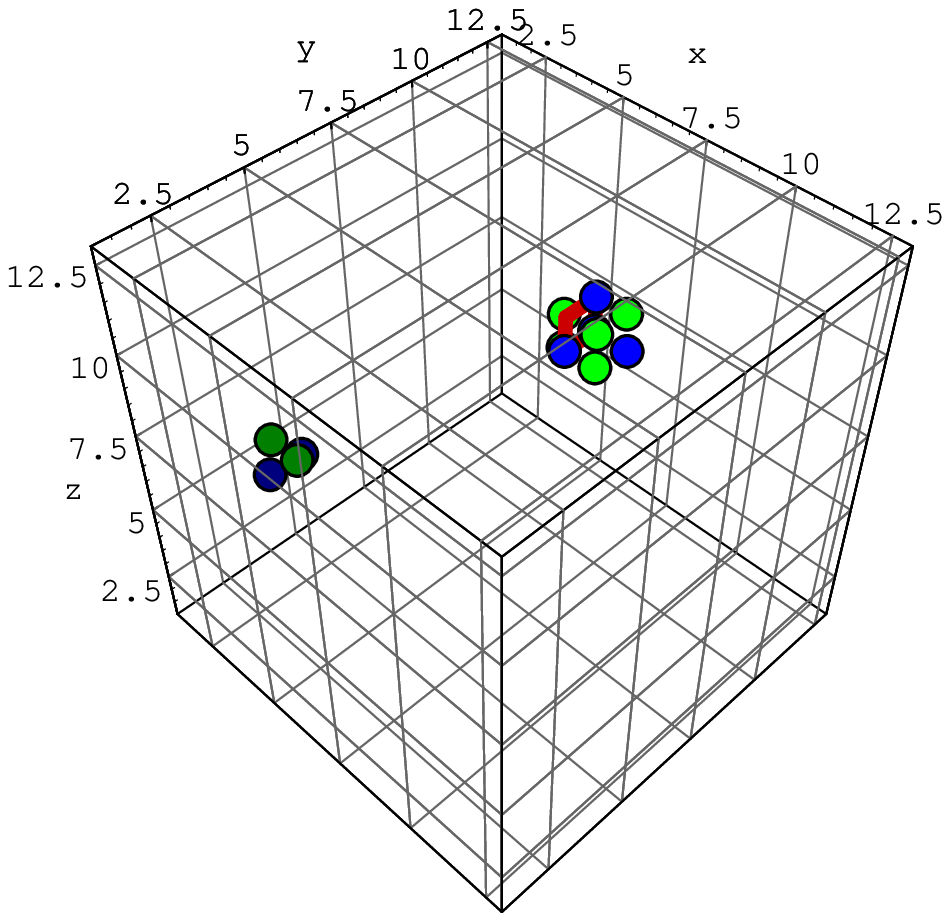}
&
\epsfxsize=5.0cm\epsffile{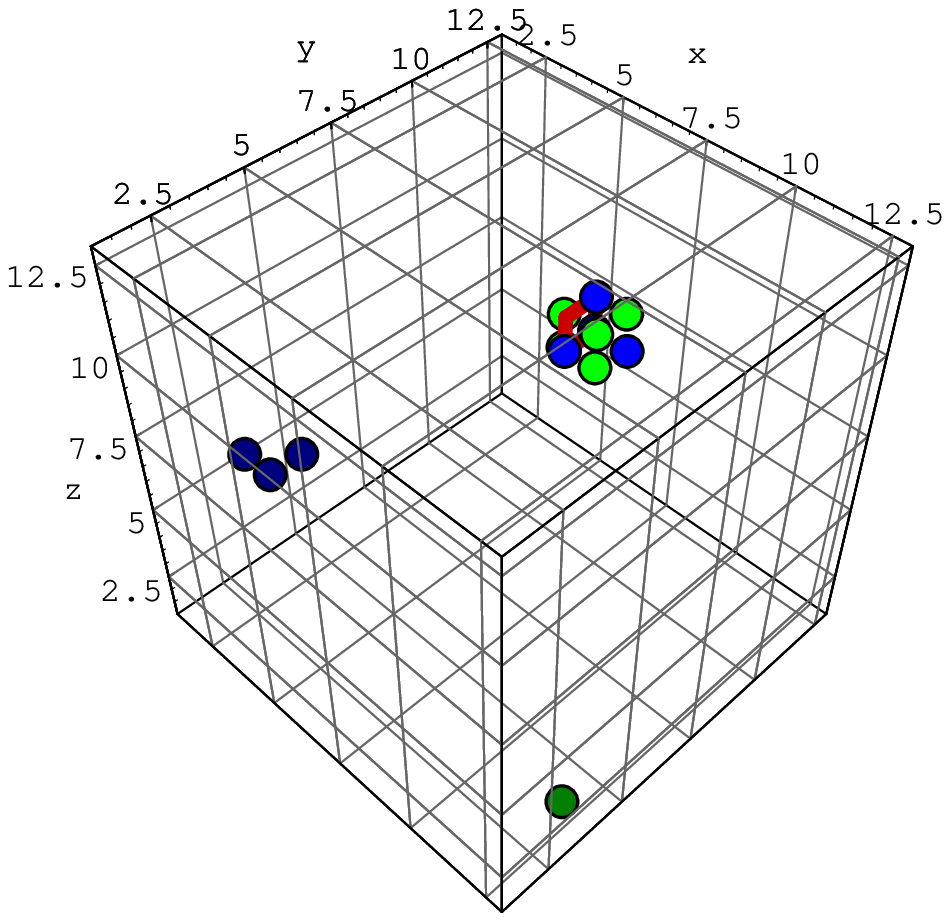}
&
\epsfxsize=5.0cm\epsffile{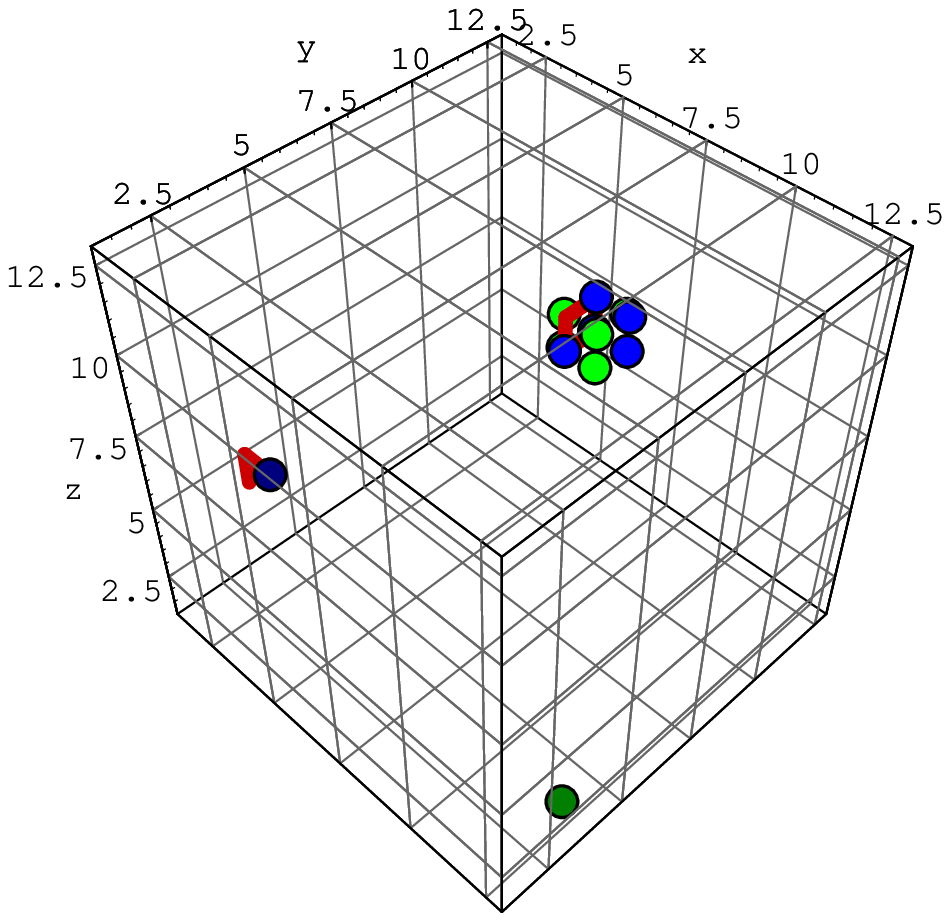}
\vspace{0.3cm}\\ 
\end{tabular}
\end{center}
{\baselineskip=13pt
Figure 4.~Cooling history  for a specific  gauge field 
configuration at a fixed time slice. The dots represent the 
topological charge distributions  in the field theoretic definitions with 
$|q(x)|>0.01$. Positive (negative) charges  are plotted with a dark (light) 
color. The monopole loops correspond to red lines. With cooling 
instantons evolve from noise accompanied by monopole loops 
in almost all cases. Note that time like monopoles cannot be seen in 
these plots. 
\baselineskip=15pt}
\end{figure*}
\clearpage
\newpage
\newpage

 
Fig. 2 shows for a fixed time slice the location of a single 
instanton (dots) which was 
put on a  trivial gauge field configuration artificially. Here a 
purely spatial monopole loop surrounds the instanton (closed line). 

In Fig. 3 we compare the results of the three methods  for the topological 
charge obtained on a single equilibrium 
gauge field after 20 cooling sweeps. From top to bottom 
the plaquette, the hypercube, and the L\"uscher definition are plotted. 
The positions of the clusters of topological charge are the same for all 
three methods. The points represent instantons or antiinstantons. 
In this particular configuration a monopole is found to 
wrap around the torus. 

Fig. 4 presents a cooling history of a time slice of a gluon field.
The topological charge using the plaquette and the
hypercube definition is displayed for cooling steps between  0 to 40. 
For any value of the plaquette
(hypercube) charge density $q(x) >
0.01$ a light blue (light green) dot was plotted. For  $q(x) <0.01$  
a dark  blue (dark green) dot was plotted. The red lines represent
the monopole loops.
Without
cooling the topological charge distribution  
cannot be identified with instantons due to noise.
Also the monopole loops do not exhibit a  structure.
After  15-20 cooling steps one can assign instantons to 
clusters of topological charge. At cooling  sweep 21 an
instanton and an antiinstanton become visible. At cooling  steps 35-40 they
begin to approach each
other and  annihilate several  steps later (not shown).
Monopole loops also thin out with cooling, but they survive in the
presence of instantons. There is
 an enhanced probability that monopole loops are present in the vicinity of
instantons in all gauge field configurations which we have checked. 

{\it 3. Conclusion and Outlook.}
Calculating the local values of topological charges and monopole currents 
and by directly displaying them with the help of  computer graphics, 
we draw a number 
of conclusions. Perhaps the most important is that after a few cooling 
sweeps one observes clearly that instantons are accompanied by monopole 
loops. This correlation  
occurs on all (semi-classical) gauge field configurations considered.
In a cooling 
history we demonstrated how instantons evolve from fluctuating 
gauge fields and how 
they are surrounded by monopoles. 
The results presented  are in nice agreement 
with earlier studies \cite{wir}, where we computed gauge averages 
of correlation functions between topological charges  and monopoles. 
There it turned out that the correlations are rather insensitive under cooling. 
Combining this finding with that of the 3D images, we conclude that the topological charge goes hand in hand with monopoles also  in the original (uncooled) 
gauge field configurations.

In this contribution we have only dealt with the quenched case. 
Switching on dynamical fermions the correlations of the topological
objects with the chiral condensate become of interest. Preliminary 
studies of such correlation functions on gauge field average have
shown non-trivial correlations with a range of about two lattice spacings 
\cite{trovie}.
This has important consequences for our understanding of the vacuum structure
of QCD because it would mean that chiral symmetry breaking occurs in the 
region of topological objects. At present we are preparing analog 3D plots
of the chiral condensate on specific gauge field configurations in order to 
check if the appearence of a non-vanishing quark condensate coincides locally
with the positions of instantons and monopoles.

\end{document}